\documentstyle[twocolumn,prl,aps,psfig]{revtex}

\begin{document}

\draft

\twocolumn[\hsize\textwidth\columnwidth\hsize\csname
@twocolumnfalse\endcsname

\title{C$_{20}$: Fullerene, Bowl or Ring? New Results
from Coupled-Cluster Calculations}

\author{Peter R. Taylor}
\address{
San Diego Supercomputer Center,
P. O. Box 85608,
San Diego, CA 92186-9784}

\author{Eric Bylaska and John H. Weare}
\address{Department of Chemistry,
University of California, San Diego,
La Jolla, CA 92093}

\author{Ryoichi Kawai}
\address{Department of Physics,
University of Alabama at Birmingham,
Birmingham, AL 35294}

\date{\today}

\maketitle

\begin{abstract}
Contrary to recent experimental evidence suggesting that the
monocyclic ring is the most stable 20-atom carbon species, highly
accurate calculations
convincingly predict that the smallest fullerene, the dodecahedron C$_{20}$,
has the lowest energy.
A related corannulene-like bowl
is nearly degenerate in energy to the fullerene.
Thermodynamic considerations suggest that at formation
temperatures of around 700 K the bowl should be the
dominant species.
The recent application of gradient corrections to LDA which supported
the ring structure is
qualitatively in error.
\end{abstract}

\pacs{PACS number: }

]

\narrowtext

The 20-atom carbon cluster is the smallest size that
can form a closed fullerene molecule.  Since small fullerenes
have been proposed as possible intermediates to C$_{60}$ or larger fullerenes
\cite{Curl:93} and because of the relevance of this
molecule to the general question of the stability of
fullerenes, this cluster has received a great deal of attention
\cite{Bowers:91,Tomanek:91,Parasuk:91a,Slanina:92,Bernholc:92,Bowers:93,%
Jarrold:93,Raghavachari:93a}.
Formation experiments by Bowers and co-workers
\cite{Bowers:91,Bowers:93}
and by Jarrold and co-workers\cite{Jarrold:93}
have probed the structure of carbon
clusters over a wide range of sizes, including C$_{20}$.
In these experiments monocyclic rings
are dominant for sizes between 11 and 30 atoms.
There is no evidence for species with a fullerene
structure for sizes less than 32 atoms. This
suggests that rings rather than fullerenes are the most stable isomers
below $n$=32.
However, in these experiments the systems are far from equilibrium.  It
may be difficult to anneal the systems to produce the most stable
species, therefore missing the fullerene.

Many theoretical calculations of
varying sophistication have been performed to try to predict stability
\cite{Tomanek:91,Parasuk:91a,Slanina:92,Bernholc:92,Bowers:93,%
Raghavachari:93a}.
It is well known from calculations on smaller clusters, C$_2$-C$_{10}$
that in order to predict the relative energies of carbon isomers,
very high accuracy methods (few kcal/mol) are required.
On the other hand, the highest level calculations that have been
reported on the C$_{20}$ molecule are based on many-body perturbation
theory (MP2).
Prior calculations using these methods for smaller systems
have shown that for carbon clusters the convergence of
perturbation theory is slow, with reliability
of the order of 50 kcal/mol
\cite{Parasuk:91a}.
Isomerization energies in C$_{20}$
and other carbon clusters can be much less
than this (roughly 10 kcal/mol), so higher accuracy
methods are necessary
to make definitive assignments of stability.
Methods with such accuracy
are commonly applied to small molecules, but rapidly become computationally
intractable as the system sizes increases.  As we discuss in this letter,
they are just feasible for the
C$_{20}$ molecule.

\begin{figure}[p]
\centerline{ \psfig{figure=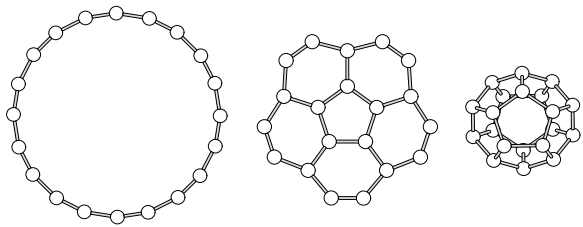,width=3in,angle=270} }
\caption{C$_{20}$ structures: ring, bowl, and fullerene)}
\label{fig:struct}
\end{figure}

Because of the expense of even low-level calculations
on systems of this size,
attention in previous work and in our calculations
has focused on three structures which appear
to be candidates for the most stable configuration, (Fig. \ref{fig:struct}):
a monocyclic ring structure (the ring I structure dominant in the
experiment by von Helden {\it et al.}\cite{Bowers:91}), a polycyclic
corannulene-like bowl comprising five fused hexagons which forms part of
the C$_{60}$ cage, and a small fullerene of near dodecahedral structure.
Some earlier estimates of the energies of these structures are summarized
in Table \ref{tab:e}.  Lowest level first-principles
calculations (Hartree-Fock self-consistent field, SCF)\cite{Parasuk:91a}
support experimental
observations\cite{Bowers:91} by assigning the ring as the lowest
energy form.  However, perturbational
improvements (MP2)\cite{Parasuk:91a} reverse
these predictions, giving the
fullerene structure the lowest energy.
On the other hand,
local density approximation (LDA) calculations
for C$_{20}$
support the fullerene as the lowest energy
structure\cite{Bernholc:92,Raghavachari:93a}.  However, when
gradient corrections\cite{Raghavachari:93a} are made to the LDA calculations,
the order is reversed, predicting the ring as the lowest energy
geometry in agreement with experiments.  The change
in the ring/fullerene separation on going from SCF to the
inclusion of correlation at the MP2 level is
roughly 150 kcal/mol.
A similar large energy difference is present between the
gradient corrected LDA and uncorrected LDA.
Given the uncertainties inherent in all
the computational methods,
none of the calculations
can be considered definitive\cite{Parasuk:92}.
We note that the reliability of the gradient corrections
is a question of considerable relevance since this method
has been suggested as a relatively tractable way to obtain
results within chemical accuracy (1 kcal/mol).

We present here results from theoretical calculations on C$_{20}$\ at
a much higher accuracy than previous work (of the order of a few
kcal/mol).
Since electron correlation effects can have a
very large influence on the energetics of the isomers,
we have performed coupled-cluster
calculations using the CCSD(T) method\cite{Raghavachari:89}.
This method includes
correlation effects of single and double replacements of
Hartree-Fock orbitals to infinite order, and in addition
a perturbative estimate of triple replacements.
Extensive series of calibration calculations have
established that the CCSD(T) method is the most accurate SCF-based
treatment of electron correlation that can feasibly be applied to
a molecule of this size.

The coupled-cluster method
provides an almost exact description of the electronic structure of
molecules that we are interested in here (see below
and in Parasuk and Alml\"of \cite{Parasuk:92})
provided a sufficiently flexible orbital basis is provided.
Unfortunately, the increase of the computational load with number of
orbital functions
is so large that completeness of the orbital basis
is a source of error.
The calculations here
were performed using Dunning's {\em cc\/}-pVDZ
basis\cite{Dunning:89} on each carbon atom.
This is a contracted set containing three $s$~functions, two
$p$~functions, and a $d$~function, obtained from a
(9$s$~4$p$~1$d$) primitive set.  It provides a ``double zeta''
description of the valence electrons (two functions per shell) and
a minimal description of the 1$s$ core electrons, plus a higher
angular momentum function to describe polarization and correlation
effects.  This level of basis set is the smallest capable of
yielding reliable results in correlated molecular calculations.
Studies (see e.g., Ref. \cite{Taylor:94}) have shown that this basis
set, optimized for treating correlation, performs better than
other sets of the same size.
Correlation effects involving the carbon
1$s$ electrons were not included because this is
expected to have a negligible effect on the predicted energetics.
19.5 million parameters were optimized iteratively in the
fullerene CCSD(T) calculation and another 29 {\em billion\/} were
estimated perturbatively.
The computational effort
is near the limit of
the capabilities of present day supercomputers even for a fixed geometry.
Therefore, it is not feasible to search for the optimum
structure at this level of calculation.
To define input geometries for the CCSD(T) calculations,
we have determined C$_{20}$\ structures at two levels: SCF and LDA
\cite{Kawai:91}.
The calculations were carried out on the CRAY C90 computer at SDSC
using the programs
{\sc molecule-sweden}\cite{molswe} and {\sc titan}\cite{titan}.
Some of the C$_6$ results discussed in this work were obtained
with {\sc aces~ii}\cite{aces}.

We should note some points about the symmetry of the various C$_{20}$
isomers.
The putative dodecahedral fullerene undergoes a
Jahn-Teller distortion: previous SCF optimization produced a
structure with $C_2$ symmetry, whereas the LDA optimization yields
a structure with $C_i$ symmetry.  Both optimized structures are
close to a dodecahedron.  The monocyclic ring is found to have
$C_{10h}$ symmetry, with alternating bonds and angles, in both SCF
and LDA optimizations.  For computational reasons, this was treated
within the subgroup $C_{2h}$ in the coupled-cluster calculations.
The bowl is found to have $C_{5v}$ symmetry at the SCF level, and
is very close to this symmetry at the LDA level.  Only the $C_s$
subgroup of $C_{5v}$ was used in the calculations.
All calculations were done for the singlet spin state: SCF
calculations demonstrate that all three structures have a
closed-shell ground state\cite{struc}.

Energies relative to the fullerene are
given in Table \ref{tab:e}.
In agreement with previous results\cite{Raghavachari:93a}, at the SCF
level, the ring is predicted to be the most stable isomer, by at
least 50 kcal/mol.
The bowl is intermediate between two other isomers
but closer to the ring in energy.
The situation is quite different using LDA energies, where
the fullerene is predicted to be most stable, with the bowl
closer to the fullerene.  The MP2 results
are similar to those of LDA.  Note that the bowl in all these
calculations is well separated from the fullerene.

The most important entry in Table \ref{tab:e} are the CCSD(T)
results.  These results
agree with the LDA results, also predicting the fullerene
to be more stable than the ring.
{\em However, at this level, the fullerene and
bowl are predicted to be essentially degenerate at the LDA geometry
and the bowl is predicted to be lower in energy at the
SCF geometries.}
This ordering is sensitive to the remaining error in the
correlation energy, but the near degeneracy of the bowl and the fullerene
will remain.

The CCSD(T) results
clearly predict that the fullerene and bowl are more
stable than the ring.
But we still need to assess the accuracy of the method.
Other
calculations have shown that for systems for which the Hartree-Fock determinent
dominates the configuration expansion,
CCSD(T) recovers essentially all the
correlation energy available with a given basis set.  The scaled
norm of the single replacements (${\cal T}_1$) as calculated by CCSD(T)
may be used as a guide
to the accuracy of a single determinent description
\cite{Taylor:89}.
If ${\cal T}_1$ is less than 0.02 the
CCSD(T) approach yields results whose reliability
depends almost entirely on the
accuracy of the atomic basis.
In these calculations,
${\cal T}_1$ was found to be 0.0153, 0.0136, and 0.0151 for the
fullerene, bowl and the ring respectively.  Thus, in a complete basis,
the accuracy will be of the
order of a few kcal/mol.  However, the {\em cc\/}-pVDZ basis
used is still quite far from complete.
For example, it
recovers only about 75\% of the valence-shell correlation energy
of the carbon atom.
Presumably, since we are looking for energy differences, much of this
error would be shared by the three structures and would cancel
in the energy difference calculation, but it is desirable to
provide some calibration for this.
The present calculations have the
largest basis that we could conceivably use in a CCSD(T) calculation
for C$_{20}$, so we can hardly perform the calibration on this
system.  Instead,
in order to ascertain the accuracy of the results
we have performed similar calculations on
the smaller system $C_6$, which we can treat with
higher accuracy basis sets.  The principal problem is the slow
convergence of the dynamical
correlation energy (that treated accurately by CCSD(T))
with angular quantum number~$l$ of the orbital basis: the behavior is
as $l^{-4}$.  Our {\em cc\/}-pDVZ
basis includes up to $d$ atomic orbitals only.
Including higher terms such as $f$ and $g$ functions increases the
computational
load very quickly.  Nevertheless,  for the C$_6$
molecule it is possible to explore basis sets up to $g$~functions
(using again Dunning's correlation-consistent
sets)\cite{Dunning:89}.  The results are summarized in
Table~\ref{tab:c6}.  More detailed results will be presented
elsewhere.\cite{Bylaska:94}
For C$_6$ we have examined three structures: a linear triplet
state, a $D_{6h}$ ring and a $D_{3h}$ ring (lowest energy).
The {\em relative\/} energies of the two
ring isomers do not change with increasing basis.  However, the
relative energy of the linear structure changes by a few kcal/mol
as the basis is increased.

The first thing to note is that the corrections in Table \ref{tab:c6}
are of the order of a few kcal/mol. Similar behavior with basis set
has been observed in $C_{10}$.\cite{Watts:92}  From this we can
confidently assert that increasing the size of the basis used for
C$_{20}$ certainly would be unlikely to change the order of the
fullerene vs. the ring structures, since this is already at least 40 kcal/mol.

The relative stability of the bowl and the fullerene is a more
difficult question.
At the present level CCSD(T) calculation, there is a considerable
difference between the correlation energy
for the different structures
($-$2.773~$E_h$ for the fullerene,
$-$2.699 for the bowl and $-$2.629 for the ring, at the SCF
geometries).
Since the main error in the calculation is in the correlation energy,
this implies that there is probably a larger energy
error in the fullerene calculation
than in the bowl or the ring.

Recently, it has been suggested by Martin\cite{Martin:94b}
that the angular basis dependence of the energy
can be extrapolated to convergence by a three-term extrapolation
formula based on the broad characteristics of the occupied orbitals
for a particular molecule.
Assuming that the majority of the correlation energy available
with a given basis for a particular system is recovered by a
CCSD(T) calculation,
the additional basis set corrections can
be estimated by first identifying the types of orbitals, i.e.
$\sigma$, $\pi$ or lone pairs, and then making corrections
which have been established for each level of basis set by comparison
with accurately measured properties for a test set of molecules.
This scheme is qualitatively consistent with the present
results.  The results in Table~\ref{tab:c6} show that the isomerization energy
from the $D_{3h}$ to $D_{6h}$ ring changes very little
with basis, because the orbital character in the two molecules
is basically the same.  On the other hand, there are major changes
in the number of $\sigma$ and $\pi$ bonds in going from the linear
to the cyclic forms, giving a significant correction in the Martin approach.
The direction of this calculated correction is consistent
with the changes observed by explicitly expanding the basis sets
used (see Table~\ref{tab:c6}).  Unfortunately we cannot
calculate the isomerization energy between the linear and ring
structures with the very largest basis set, because the
calculation on the (triplet spin state) linear form is not feasible
with this large a basis.

As an additional check on the
Martin approach, we can refer to the total atomization energy of
the C$_3$ molecule using the {\em cc\/}-pDVZ basis (see also
Ref.~\cite{Martin:94a}).  In this case
a 30kcal/mol basis set
error in the atomization energy is predicted by the Martin
formula: applying this correction brings the predicted atomization
energy within the error bars of the measured value.

The stabilities of the C$_{20}$ isomers estimated via the Martin
approach are included as
the last line in Table~\ref{tab:e}.  For both the SCF and the
LDA geometries, the fullerene is now more stable than the bowl.
We note, however, that at the SCF geometries the separation is
only 5 kcal/mol, which is roughly the accuracy of the calculation.
Therefore, these calculations only weakly support the fullerene as
the most stable form.

Another source of error is the lack of CCSD(T) geometry
optimization for
each isomer.  In the calculations, the ring and the bowl had lower
CCSD(T) energies
at the SCF optimized geometries, while the fullerene was lower at
the LDA geometry.  This is naively consistent with the greater
importance of electron correlation in the fullerene, compared to
the other isomers.  It is difficult to say with any certainty what
would be the results of a CCSD(T) geometry optimization, but if we
accept the proposition that correlation affects the fullerene more
than the other isomers, we would expect geometry optimization to
further stabilize the fullerene relative to the other isomers.

The relative stabilities we have presented are all for
hypothetical vibrationless states at absolute zero.  The zero-point
vibrational energies for the three isomers are 69.8, 73.3, and
76.8~kcal/mol for ring, bowl, and fullerene, respectively, using
published SCF frequencies~\cite{Raghavachari:93a}.
Still the fullerene is most stable.
However, experimental data is taken at high temperature. Therefore,
it is desirable to explore finite-temperature effects.
While there is no doubt that these effects
will favour the ring isomer due to its
higher vibrational entropy , the
relative energy of the ring versus the fullerene is so large that
these effects alone cannot explain the nonappearance of the
fullerene in a fully equilibrated system at
temperatures less than 1500 K
(estimated from the SCF frequencies).
On the other hand, even with extrapolations
the fullerene and the bowl are quite close in energy.
Using an average separation of
9kcal/mol between the bowl and the fullerene, we estimate that above
approximately 700 K
the bowl is more stable.  Below these temperatures
the fullerene will be progressively more stable.  However,
it is easy to believe
that the transition state between any of these structures is very high
and dynamically difficult to achieve,
making it very unlikely for the transition from fullerene to
bowl to occur.  This is
supported by our dynamical calculations which show these molecules
to be stable even at very high temperatures.

Finally, our CCSD(T) calculations clearly show that
a gradient correction to LDA, at least as implemented in
Raghavachari {\it et al.}, is incorrect (by 165~kcal/mol!) for this problem.
This is unfortunate,
since the gradient correction approach has been widely
advertised as a solution to the well-known
energetics problems of LDA, and is relatively easy to implement.
We note, however, that the uncorrected LDA gives predictions
that are semiquantitatively correct and
much better than SCF for a similar computational cost.

This work was supported in part by National Science Foundation
Cooperative Agreement DASC-8902825, by Office of Naval
Research (grant number NOO14-91-J-1835), and by a grant of
computer time from the San Diego Supercomputer Center.  We would
particularly like to acknowledge the contributions of
C. W. Bauschlicher, T. J. Lee, and A. P. Rendell to {\sc
molecule-sweden\/} and {\sc titan\/} that made these calculations
possible, and the assistance of M.~Vildibill in running the
C$_{20}$ calculations.


\begin{thebibliography}{99}
\bibitem{struc}
The LDA and SCF optimized geometries, energies, etc., can be
obtained by anonymous FTP from the site ftp.sdsc.edu, in the
directory /pub/sdsc/chemistry/structures/C20.
\end{thebibliography}

\begin{table}[t]
\begin{center}
\caption{C$_{20}$\ energies (eV) relative to the fullerene}
\label{tab:e}
\vspace{7 pt}
\begin{tabular}{lcrrrcrr}
& \multicolumn{3}{c}{LDA geometries} &
& \multicolumn{3}{c}{SCF geometries} \\
Method & fullerene & bowl & \hspace{0.1in}ring
&\hspace{0.2in} & fullerene & bowl & \hspace{0.1in}ring \\
\hline
SCF$^a$   & & & & & 0. & --- & $-$4.6 \\
MP2$^a$   & & & & & 0. & --- & 3.7 \\
LDA$^b$   & & & & & 0. & 1.0 & 3.8  \\
GC-LDA$^c$   & & & & & 0. & $-$2.4 & $-$3.4  \\
SCF & 0. & $-$2.0 & $-$2.2 & & 0. & $-$2.5 & $-$3.4 \\
LDA & 0. & 0.8    & 2.3    & & 0. & 0.8    & 3.3    \\
MP2 & 0. & 1.1 & 2.8 & & 0. & 0.8 & 3.9 \\
CCSD(T) & 0. & 0.0 & 1.7 & & 0. & $-$0.6 & 2.2 \\
estimate$^c$ & 0. & 0.7 & 2.4 & & 0. & 0.2 & 2.6
\end{tabular}
\end{center}
\noindent
$^a$ Parasuk and Almlof\cite{Parasuk:91a}

\noindent
$^b$ Raghavachari {\it et al.}\cite{Raghavachari:93a}

\noindent
$^c$ Gradient corrected LDA \cite{Raghavachari:93a}

\noindent
$^d$ See text.
\end{table}

\begin{table}
\begin{center}
\caption{C$_6$ Convergence of isomerization energies with basis set
size. (kcal/mol)}
\label{tab:c6}
\vspace{7pt}
\begin{tabular}{lrrrr}
&spd&spdf&spdfg&estimate \\ \hline
$D_{3h}$&0.00&0.00&0.00&0.00 \\
$D_{6h}$&8.2&8.2&7.9&8. $\pm$ .2  \\
linear&5.0&9.4&---&14. $\pm$ 1.  \\
\end{tabular}
\end{center}
\end{table}

\end{document}